\documentclass[preprintnumbers,amsmath,amssymb,eqsecnum]{revtex4}

\usepackage{epsf}
\usepackage{graphicx}  
\usepackage{dcolumn}   
\usepackage{bm}        

\begin{document}

\preprint{hep-th/0410158}

\title{Born-Infeld Black Holes in (A)dS Spaces}

 \author{Rong-Gen Cai\footnote{e-mail address:
cairg@itp.ac.cn}}

\address{CASPER, Department of Physics, Baylor University, Waco,
 TX76798-7316, USA \\
  Institute of Theoretical Physics, Chinese
Academy of Sciences,
 P.O. Box 2735, Beijing 100080, China}

\author{Da-Wei Pang\footnote{e-mail address:
pangdw@itp.ac.cn}}
\address{Institute of Theoretical Physics, Chinese
Academy of Sciences,
 P.O. Box 2735, Beijing 100080, China\\
 and Graduate School of the Chinese Academy of Sciences}

\author{Anzhong Wang\footnote{e-mail address:
Anzhong\_Wang@baylor.edu}}

\address{CASPER, Department of Physics, Baylor University, Waco,
 TX76798-7316, USA}

\begin{abstract}
We study some exact solutions in a $D(\ge4)$-dimensional
Einstein-Born-Infeld theory with a cosmological constant. These
solutions are asymptotically de Sitter or anti-de Sitter,
depending on the sign of the cosmological constant. Black hole
horizon and cosmological horizon in these spacetimes can be a
positive, zero or negative constant curvature hypersurface. We
discuss the thermodynamics associated with black hole horizon and
cosmological horizon.  In particular we find that for the
Born-Infeld black holes with Ricci flat or hyperbolic horizon in
AdS space, they are always thermodynamically stable, and that for
the case with a positive constant curvature, there is a critical
value for the Born-Infeld parameter, above which the black hole is
also always thermodynamically stable, and below which a unstable
black hole phase appears.  In addition, we show that although the
Born-Infeld electrodynamics is non-linear, both black hole horizon
entropy and cosmological horizon entropy can be expressed in terms
of the Cardy-Verlinde formula. We also find a factorized solution
in the Einstein-Born-Infeld theory, which is a direct product of
two constant curvature spaces: one is a two-dimensional de Sitter
or anti-de Sitter space, the other is a ($D-2$)-dimensional
positive, zero or negative constant curvature space.

\end{abstract}
\maketitle

\section{Introduction}

Black holes in anti-de Sitter (AdS) and de Sitter (dS) spaces are
quite different from their counterparts in asymptotically flat
spaces. In AdS spaces, there are so-called topological black holes
whose event horizon could be a positive, zero or negative constant
curvature surface~\cite{topo}. In asymptotically dS spaces, there
are no globally time-like Killing vector and spatial infinity,
there is still not a well defined approach to define conserved
charges like mass and angular momentum of asymptotically dS
gravitational configurations~\cite{ab,bbm,CMZ}. Due to the AdS/CFT
correspondence~\cite{AdS}, it was argued by Witten~\cite{witten}
that the thermodynamics of black holes in AdS spaces can be
identified with that of dual conformal field theory (CFT) residing
on the boundary of the AdS space. In the sense of the dS/CFT
correspondence~\cite{Stro}, thermodynamics of black hole horizon
and of cosmological horizon in the asymptotically de Sitter spaces
might be related to that of dual CFTs. These motivate a surge of
study of black holes in AdS and dS spaces.

For a $(1+1)$-dimensional conformal field theory, there is a
well-known entropy formula, namely, the Cardy
formula~\cite{Cardy}. In an elegant paper~\cite{Ver}, in the
spirit of AdS/CFT correspondence, Verlinde argued that there is a
similar entropy formula  for CFTs in higher dimensions. The
formula ``derived" by Verlinde is called Cardy-Verlinde formula in
the literature. Indeed  this formula
 has been checked to hold for CFTs with
AdS gravity duals, such as Schwarzschild-AdS black
holes~\cite{Ver}, Kerr-AdS black holes~\cite{KPS}, Hyperbolic and
charged black holes~\cite{Cai}, Taub-Bolt-AdS
instanton~\cite{Birm}, Kerr-Newmann-AdS black holes~\cite{Jing}
and so on (see also \cite{Muck}). In addition, it has been found
that entropies of black hole horizons~\cite{Cai2} and cosmological
horizons~\cite{Cai3} in asymptotically dS spaces can also be
expressed in terms of the Cardy-Verlinde formula.

In 1934 Born and Infeld proposed a non-linear electrodynamics with
the aim of obtaining a finite value for the self-energy of a
point-like charge~\cite{BI}. Although it became less popular with
the introduction of QED, in recent years the Born-Infeld action
has been occurring repeatedly with the development of superstring
theory, where the dynamics of D-branes, some soliton solutions of
supergravity, is governed by the Born-Infeld action. For various
motivations, extending the Reissner-Nordstr\"om black hole
solutions in Einstein-Maxwell theory to the charged black hole
solutions in Einstein-Born-Infeld theory with/without a
cosmological constant has  attracted some attention in recent
years, for example, see \cite{BIBH,Dey}.

In this paper, we first generalize the exact solutions of
spherically symmetric Born-Infeld black holes with a cosmological
horizon in arbitrary dimensions recently given in \cite{Dey} to
the case where black hole horizon and/or cosmological horizon is a
positive, zero or negative constant curvature surface. We then in
Secs.~III and IV  study thermodynamics associated with black hole
horizon and cosmological horizon and show that although the
electrodynamics is non-linear, entropies of black holes and
cosmological horizon in the Einstein-Born-Infeld theory can still
 be expressed in terms of the Cardy-Verlinde formula, which shows
the latter is of some universality. In Sec.~V we present a
factorized solution in the Einstein-Born-Infeld theory. The
solution is a direct product of two constant curvature spaces: one
is a two-dimensional dS or AdS space and the other is a positive,
zero or negative constant curvature space. The result is
summarized in Sec.~VI.

\section{Born-Infeld black holes in (A)dS spaces}

Consider an $(n+1)$-dimensional ($n \ge 3$) Einstein-Born-Infeld
action with a cosmological constant $\Lambda$
\begin{equation}
S=\int d^{n+1}x\sqrt{-g}\left(\frac{{\cal R}-2\Lambda}{16\pi
G}+L(F)\right),
\end{equation}
where ${\cal R}$ is scalar curvature and $L(F)$ is given by
\begin{equation}
L(F)=4\beta^{2}\left(1-\sqrt{1+\frac{F^{\mu\nu}F_{\mu\nu}}{2\beta^{2}}}\right
).
\end{equation}
Here the constant $\beta $ is called Born-Infeld parameter with
dimension of mass. In the limit $\beta\rightarrow\infty$, $L(F)$
reduces to the standard Maxwell form
\begin{equation}
L(F)=-F^{\mu\nu}F_{\mu\nu}+{\cal O}(F^{4}),
\end{equation}
while one has $L(F)\to 0$ as $\beta \to 0$. In what follows we set
$16\pi G=1$ for simplicity, where $G$ is the Newton constant in
$(n+1)$-dimensions.

The equations of motion of the electromagnetic field and the
Einstein equations can be obtained by varying the action with
respect to the gauge field $A_{\mu}$  and the metric $g_{\mu\nu}$,
which yields
\begin{equation}
\label{2eq4}
\partial_{\mu}\left
 (\frac{\sqrt{-g}F^{\mu\nu}}{\sqrt{1+\frac{F^{\mu\nu}F_{\mu\nu}}{2\beta^{2}}}}\right
 )=0,
\end{equation}
\begin{equation}
\label{2eq5}
 {\cal R}_{\mu\nu}-\frac{1}{2}{\cal R}g_{\mu\nu}+\Lambda
g_{\mu\nu}=\frac{1}{2}g_{\mu\nu}L(F)+ \frac{2F_{\mu
\alpha}F^{~\alpha}_{\nu}}{\sqrt{1+\frac{F^{\mu\nu}F_{\mu\nu}}{2\beta^{2}}}},
\end{equation}
respectively. Here ${\cal R}_{\mu\nu}$ stands for Ricci tensor.

Suppose the spacetime metric is of the form
\begin{equation}
\label{2eq6}
ds^{2}=-V(r)dt^{2}+\frac{1}{V(r)}dr^{2}+R^{2}(r)h_{ij}dx^{i}dx^{j},
\end{equation}
where $V(r)$ and $R(r)$ are two functions of the coordinate $r$
only, and  $h_{ij}$ is a function of coordinates $x^{i}$ which
span an $(n-1)$-dimensional hypersurface with constant scalar
curvature  $(n-1)(n-2)k$. Here $k$ is a constant and characterizes
the hypersurface. Without loss of generality, one can take $k$ to
be $\pm 1$ and $0$ such that the black hole horizon or
cosmological horizon in (\ref{2eq6}) can be a positive (elliptic),
zero (flat) or negative (hyperbolic) constant curvature
hypersurface. For the metric (\ref{2eq6}), we have non-vanishing
components of Ricci tensor
\begin{eqnarray}
\label{2eq7}
 && {\cal R}^{t}_{t}= -\frac{V''}{2} - (n-1)\frac{V'R'}{2R},
   \\
   \label{2eq8}
&& {\cal R}^{r}_{r}=
-\frac{V''}{2}-(n-1)\frac{V'R'}{2R}-(n-1)\frac{VR''}{R},
\\
\label{2eq9}
 && {\cal R}^{i}_{j}= \left( \frac{n-2}{R^2}k
-\frac{1}{(n-1)R^{n-1}}[V(R^{n-1})']'\right)\delta^{i}_{j},
\end{eqnarray}
where a prime stands for the derivative with respect to the
coordinate $r$.

In the static and symmetric background (\ref{2eq6}), the equation
(\ref{2eq4}) can be satisfied  by setting $F^{\mu\nu}$ to zero,
except for $F^{rt}$, which gives
\begin{equation}
\label{2eq10}
 F^{rt}=\frac{\sqrt{(n-1)(n-2)}\beta
q}{\sqrt{2\beta^{2}R^{2n-2}+(n-1)(n-2)q^{2}}},
\end{equation}
where $q$ is an integration constant relating to the electric
charge of the solution. Defining the electric charge via $Q=
\frac{1}{4\pi} \int \ ^*F d\Omega$, we have
\begin{equation}
\label{2eq11}
 Q = \frac{\sqrt{(n-1)(n-2)}\omega_{n-1}}
{4\pi\sqrt{2}}q,
\end{equation}
where $\omega_{n-1}$ represents the volume of constant curvature
hypersurface described by $h_{ij}dx^idx^j$.

From Eqs. (\ref{2eq7}) and (\ref{2eq8}), we have $R''(r)=0$, which
has two solutions. One is $R=r$; the other is $R=a$, here $a$ is a
constant. We first consider the case of $R=r$. The case of $R=a$
will be discussed later. With $R=r$, solving the equation
(\ref{2eq9}) yields
\begin{eqnarray}
\label{2eq12}
 V(r) &=&  k -\frac{m}{r^{n-2}} +\left(
\frac{4\beta^2}{n(n-1)} +\frac{1}{l^2}\right) r^2
 \nonumber \\
&& -\frac{2\sqrt{2} \beta}{(n-1)r^{n-2}}\int
\sqrt{2\beta^2r^{2n-2} +(n-1)(n-2)q^2 }dr.
\end{eqnarray}
Here we have redefined the cosmological constant as $\Lambda =
-n(n-1)/2l^2$. It turns out that the integration in (\ref{2eq12})
can be worked out and can be expressed by using a hypergeometric
function,
\begin{eqnarray}
\label{2eq13}
V(r) &=&k-\frac{m}{r^{n-2}}+\left(\frac{4\beta^{2}}{n(n-1)}+\frac{1}{l^{2}}\right) r^{2}
    \nonumber \\
    &&-\frac{2\sqrt{2}\beta}{n(n-1)r^{n-3}}\sqrt{2\beta^{2}r^{2n-2}+(n-1)(n-2)q^{2}}
      \nonumber \\
    &&+\frac{2(n-1)q^{2}}{nr^{2n-4} } \ _2F_{1}[\frac{n-2}{2(n-1)},\frac{1}{2},
    \frac{3n-4}{2(n-1)},
    -\frac{(n-1)(n-2)q^{2}}{2\beta^{2}r^{2n-2}}].
\end{eqnarray}
Here $m$ is an integration constant, relating to the mass of the
gravitational configuration. Since the metric is asymptotically de
Sitter ($l^2<0$) or anti-de Sitter ($l^2>0$), according to the
definition of mass in asymptotically dS and AdS spaces due to
Abbott and Deser~\cite{ab}, we have the mass of the solution
\begin{equation}
\label{2eq14}
M =(n-1)\omega_{n-1}m.
\end{equation}
Note that throughout this paper, the convention $16\pi G=1$ has
been used. When $k=1$ and $h_{ij}dx^idx^j$ denotes the line
element of an $(n-1)$-dimensional unit round sphere, the solution
(\ref{2eq13}) reduces to the one found in \cite{Dey}. Note that
for a positive constant curvature space, the metric is not
necessary to be a round sphere. Therefore the metric (\ref{2eq6})
is  more general than a spherically symmetric metric.

\section{Cardy-Verlinde formula for the Born-Infeld Black Holes in
AdS spaces}

In this section we first consider the case of $l^2>0$, namely for
a negative cosmological constant. In this case, the spacetime
 asymptotically approaches to an AdS space. The solution (\ref{2eq13})
 describes a Born-Infeld black hole in AdS space. The black hole horizon
 is determined by $V(r)|_{r=r_+}=0$. The behavior of metric function $V$ in
the small $r$ region and large $r$ region and thermodynamics of
the black hole for the case of $k=1$ have been analyzed and
discussed in \cite{Dey}. We will not therefore repeat them here,
instead we will mainly focus on the cases of $k=0$ and $k=-1$ and
will show that the black hole entropy can be expressed in terms of
the Cardy-Verlinde formula.

The temperature of the black hole can be obtained by continuing
the metric (\ref{2eq6}) to its Euclidean sector via $t=-i\tau$ and
requiring the absence of conical singularity at the horizon. This
results in a periodic Euclidean time $\tau$ with period $1/T$,
which is just the inverse Hawking temperature of the black hole.
Calculation gives
\begin{equation}
\label{3eq1}
 T=\frac{1}{4\pi r_+}
 \left ((n-2)k+\left
 (\frac{4\beta^{2}}{n-1}+\frac{n}{l^{2}}\right )r_{+}^2
  -\frac{2\sqrt{2}\beta}{(n-1)r^{n-3}_{+}}\sqrt{2\beta^{2}r^{2n-2}_{+}+(n-1)(n-2)q^{2}}
  \right ).
\end{equation}
When $T=0$, the black hole is an extremal black hole, there the
black hole horizon and the internal horizon (Cauchy horizon)
coincide with each other. For the extremal black hole, the charge
has a relation to the horizon radius
 \begin{equation}
 \label{3eq2}
 (n-1)(n-2) q^2 =
 -2\beta^2r_+^{2n-2}+\frac{(n-1)^2r_+^{2n-6}}{8\beta^2}
  \left ((n-2)k
  +\left(\frac{4\beta^2}{n-1}+\frac{n}{l^2}\right)r_+^2
   \right)^2.
\end{equation}
 The entropy
of the black hole still obeys the so-called horizon area formula
\begin{equation}
\label{3eq3}
 S=4\pi\omega_{n-1}r^{n-1}_{+}.
\end{equation}
It is easy to show that these thermodynamic quantities such as
charge (\ref{2eq11}), mass (\ref{2eq14}), temperature (\ref{3eq1})
and entropy (\ref{3eq3}) satisfy the first law of black hole
thermodynamics
\begin{equation}
\label{3eq4}
 dM=TdS+\Phi dQ,
\end{equation}
where $\Phi$ is the electrostatic potential at the black hole
horizon, which is conjugate to the electric charge $Q$,
\begin{equation}
\label{3eq5}
 \Phi=\sqrt{\frac{n-1}{2(n-2)}}\frac{16\pi
 q}{r^{n-2}_{+}} \ _{2}F_{1}[\frac{n-2}{2(n-1)},
\frac{1}{2},\frac{3n-4}{2(n-1)},-\frac{(n-1)(n-2)q^{2}}{2\beta^{2}r^{2n-2}_{+}}].
\end{equation}
Note that the mass of black hole can be expressed in terms of the
horizon radius $r_+$
\begin{eqnarray}
\label{3eq6} M &=& (n-1)\omega_{n-1} r_+^{n-2} \left (k +\left
(\frac{4\beta^{2}}{n(n-1)}+\frac{1}{l^{2}}
   \right) r^{2}_{+} \right. \nonumber \\
   && -\frac{2\sqrt{2}\beta }{n(n-1)r_+^{n-3}}\sqrt{2\beta^{2}r^{2n-2}_{+}+(n-1)(n-2)q^{2}}
    \nonumber \\
   && \left . +\frac{2(n-1)q^{2}}{nr^{2n-4}_{+}} \ _{2}F_{1}[\frac{n-2}{2(n-1)},\frac{1}{2},
   \frac{3n-4}{2(n-1)},
   -\frac{(n-1)(n-2)q^{2}}{2\beta^{2}r^{2n-2}_{+}}]\right ).
\end{eqnarray}

To see the thermodynamical stability of the black hole, let us
calculate the heat capacity with a fixed charge, which implies
that we are discussing the stability in a canonical ensemble.
\begin{eqnarray}
\label{3eq7}
 C_q &=& \left(\frac{\partial M}{\partial T}\right)_q \nonumber \\
     &=& \frac{16\pi^2 r_+^nT}{-(n-2)k +\left
     (\frac{4\beta^2}{n-1} +\frac{n}{l^2}\right)r_+^2 +
     \frac{2\sqrt{2}\beta}{(n-1)r_+^{n-3}}\frac{-2\beta^2r_+^{2n-2}
       +(n-1)(n-2)^2
       q^2}{\sqrt{2\beta^2r_+^{2n-2}+(n-1)(n-2)q^2}}}.
\end{eqnarray}
Note that from the metric (\ref{2eq13}), Hawking temperature
(\ref{3eq1}) or the heat capacity (\ref{3eq7}), one can see that
the case of $\beta=0$ or $q=0$ reduces to the case of
Schwarzschild black hole in AdS space. In other words, when $\beta
=0$, the charge parameter $q$ will disappear automatically, and
vice versa. When $q=0$, we can see that the heat capacity is
always positive if $k=0$ or $-1$, while it is negative for $r_+
<l\sqrt{(n-2)/n}$, positive for $r_+
> l\sqrt{(n-2)/n}$, and diverges at $r_+=l\sqrt{(n-2)/n}$, when
$k=1$. When $q \ne 0$, since the numerator in (\ref{3eq7}) is
always positive, the sign of $C_q$ is therefore completely
determined by that of the denominator, from which we are certainly
able to get a relation of the horizon radius to other parameters
$k$, $\beta$, spacetime dimension $n$ and charge $q$. Numerical
check indicates that when $k=0$ or $-1$, the heat capacity is
always positive. This can be seen from the behavior of Hawking
temperature (\ref{3eq1}). In Fig.~1 and Fig.~2, we respectively
plot the inverse Hawking temperature for a given charge when $k=0$
and $k=-1$. We see that the inverse Hawking temperature always
monotonically decreases from infinity (corresponding to extremal
black holes) to zero as the black hole horizon radius increases.
Note that the region with a negative Hawking temperature should be
excluded from the physical phase space.
\begin{figure}[ht]
\includegraphics[totalheight=1.7in]{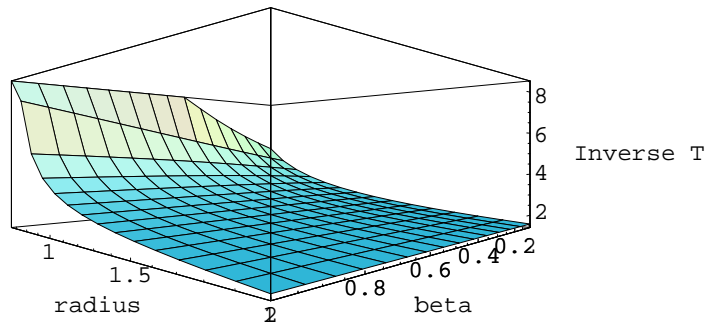}
 \caption{The inverse Hawking temperature of a five-dimensional Born-Infeld black
 hole in AdS space for a fixed charge $q/l^2 =0.5$ for the case of $k=0$.}
\end{figure}

\begin{figure}[ht]
\includegraphics[totalheight=1.7in]{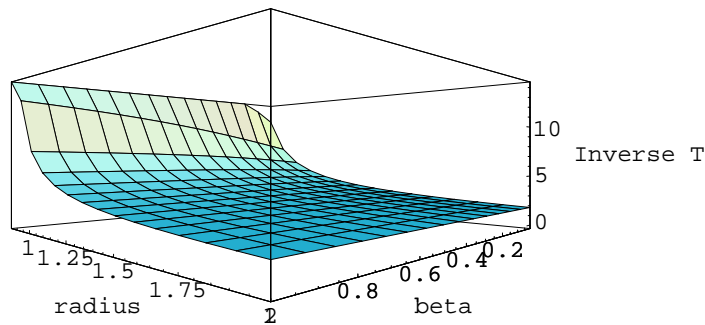}
 \caption{The inverse Hawking temperature of a five-dimensional Born-Infeld black
 hole in AdS space for a fixed charge $q/l^2 =0.5$ for the case of $k=-1$.}
\end{figure}

When $k=1$, the heat capacity can be positive or negative.  We
plot the inverse Hawking temperature of the Born-Infeld black hole
when $k=1$ in Fig.~3. In the case of $k=1$, we notice that between
two stable  black hole phases (small black holes near the extremal
limit and large black holes where the effect of charge is
negligible) there is a unstable black hole phase. However, this
unstable black hole phase will disappear when the Born-Infeld
parameter $\beta$ increases. To clearly see this point, in Fig.~4
we plot the inverse temperatures of black holes with a fixed
charge, but with different $\beta$, from which we can see clearly
how the unstable phase disappears. This implies that for the
Born-Infeld black holes with a given charge, there is a critical
value of the Born-Infeld parameter $\beta$, above which the black
holes are always stable thermodynamically. This critical point can
be calculated from the Hawking temperature (\ref{3eq1}). The
expression, however, is a little bit complicated, we do not
therefore present it here.

\begin{figure}[ht]
\includegraphics[totalheight=1.7in]{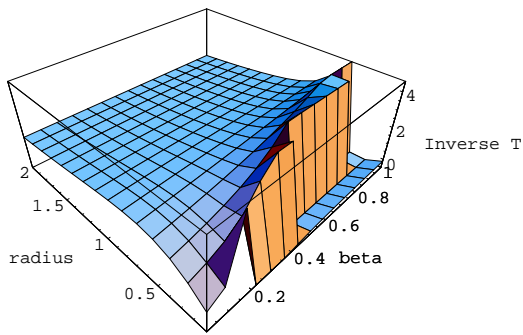}
 \caption{The inverse Hawking temperature of a five-dimensional Born-Infeld black
 hole in AdS space for a fixed charge $q/l^2 =0.5$ for the case of $k=1$.}
\end{figure}

\begin{figure}[ht]
\includegraphics[totalheight=1.7in]{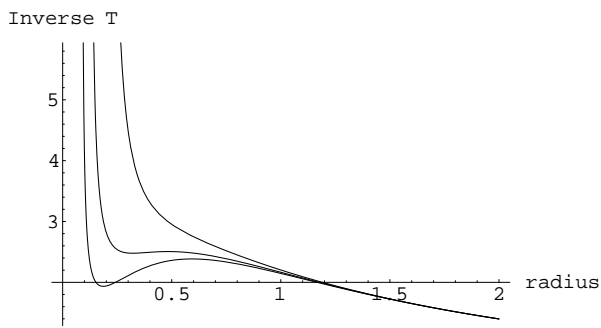}
 \caption{The inverse Hawking temperature of a five-dimensional Born-Infeld black
 hole in AdS space for a fixed charge $q/l^2 =0.5$ for the case of $k=1$. Three curves
 from right to left correspond to the case $\beta l =0.40$, $0.22$ and $0.15$.}
\end{figure}

According to the AdS/CFT correspondence, the thermodynamics of
Born-Infeld black holes in AdS spaces should also be identified
with that of a some dual CFT. Since the entropy of CFTs with AdS
duals can be described by the Cardy-Verlinde formula~\cite{Ver},
so it is of interest to see whether or not the black hole entropy
(\ref{3eq3}) can be recast to a form of the Cardy-Verlinde
formula.

Suppose there is a CFT residing on an ($n-1$)-dimensional sphere
with radius $R$,
\begin{equation}
ds^{2}=-dt^{2}+R^{2}d\Omega^{2}_{n-1}.
\end{equation}
The Cardy-Verlinde formula can be written down as~\cite{Ver}
\begin{equation}
S=\frac{2\pi R}{n-1}\sqrt{E_{c}(2E-E_{c})},
\end{equation}
where $E$ and $E_{c}$  respectively represent the total energy and
the Casimir energy, non-extensive part of energy, of the CFT.

For the Born-Infeld black holes, the total energy $E$ is the AD
mass $M$ given in (\ref{2eq14}). The corresponding Casimir energy
$E_{c}$ is~\cite{Ver,Cai}
\begin{equation}
\label{3eq10}
 E_{c}=nE-(n-1)TS-(n-1)\Phi Q.
\end{equation}
Substituting those thermodynamic quantities into (\ref{3eq10}), we
get
\begin{equation}
\label{3eq11}
 E_{c}=2k(n-1)r^{n-2}_{+}\omega_{n-1}.
\end{equation}
We see that for Born-Infeld black holes with Ricci flat horizon
($k=0$), the Casimir energy vanishes, while it is negative for the
hyperbolic horizon.

With the energy $E$ and the Casimir energy $E_c$, we find that the
black hole entropy (\ref{3eq3}) can indeed be cast to a form of
the Cardy-Verlinde formula as follows,
\begin{equation}
\label{3eq12}
 S=\frac{2\pi l}{n-1}\sqrt{(2(E-E_{q})-E_{c})E_{c}/k}.
\end{equation}
The dual CFT to the Born-Infeld black holes resides on the
boundary of the bulk metric (\ref{2eq6}). Up to a conformal
factor, the boundary metric is
\begin{equation}
ds^2 =- dt^2 +l^2 h_{ij}dx^idx^j,
\end{equation}
it is just the boundary spacetime, in which the dual CFT resides
and its entropy is described by the formula (\ref{3eq12}). $E_q$
in (\ref{3eq12}) is nothing but the energy of electromagnetic
field outside the black hole horizon, which can be calculated as
\begin{eqnarray}
\label{3eq14}
 E_{q} &\equiv & \omega_{n-1}\int^{\infty}_{r_{+}}\bar T^{0}_{0}r^{n-1}dr
 \nonumber \\
     &=& \omega_{n-1} \left( \frac{4\beta^{2}}{n}r^{n}_{+}-\frac{2\sqrt{2}\beta r_{+}}
     {n}\sqrt{2\beta^{2}r^{2n-2}_{+}+(n-1)(n-2)q^{2}}\right.  \nonumber \\
     && + \left. \frac{2(n-1)^{2}q^{2}}{nr^{n-2}_{+}} \ _{2}F_{1}[\frac{n-2}{2(n-1)},\frac{1}{2},
     \frac{3n-4}{2(n-1)},-\frac{(n-1)(n-2)q^{2}}{2\beta^{2}r^{2n-2}_{+}}]\right) ,
\end{eqnarray}
where  $\bar T^{0}_{0}$ is the $0-0$ component of the
energy-momentum tensor of non-linear electromagnetic field,
\begin{equation}
\bar T_{\mu\nu}=\frac{1}{2}g_{\mu\nu}L(F)+ \frac{2F_{\mu
\alpha}F^{~\alpha}_{\nu}}{\sqrt{1+\frac{F^{\mu\nu}F_{\mu\nu}}{2\beta^{2}}}}.
\end{equation}
Thus we show that although the electrodynamics in the
Einstein-Born-Infeld theory is non-linear, the entropy of the
Born-Infeld black holes in AdS space can still be expressed in
terms of the Cardy-Verlinde formula. Of course, the energy of
electromagnetic field outside the black hole horizon should be
subtracted from the total energy.

\section{Cardy-Verlinde formula for the Born-Infeld black holes in
dS spaces}

In this section we consider the Born-Infeld black holes in de
Sitter space ($l^2<0$ in (\ref{2eq12})). In this case, a
cosmological horizon appears, in addition to  black hole horizons.
Both cosmological horizon and black hole horizon have
thermodynamic properties such as Hawking temperature, entropy and
etc.  As we mentioned in Sec.~I, it is not an easy matter to
compute conserved charges associated with an asymptotically dS
space because of the absence of spatial infinity and a globally
timelike Killing vector in such a spacetime. It is found that if
one uses the definition of mass due to Abbott and Deser in dS
space~\cite{ab}, the entropy of Schwarzschild black holes in dS
space can be expressed by the Cardy-Verlinde formula, but not for
the entropy of cosmological horizon~\cite{Cai2}. On the other
hand, if one uses the definition of mass coming from the surface
counterterm prescription in asymptotically dS
spaces~\cite{bbm,CMZ}, the entropy of cosmological horizon can be
described by the Cardy-Verlinde formula, but not for the entropy
of black hole horizon~\cite{Cai3}.

For convenience, we change $l^2$ in (\ref{2eq13}) to $-l^2$ so
that in what follows one has $l^2>0$. In this case, the metric
function $V$ becomes
\begin{eqnarray}
\label{4eq1}
V(r)&=& k-\frac{m}{r^{n-2}}+\left( \frac{4\beta^{2}}{n(n-1)}-\frac{1}{l^{2}}\right )r^{2}
  \nonumber \\
    &&-\frac{2\sqrt{2}\beta}{n(n-1)r^{n-3}}\sqrt{2\beta^{2}r^{2n-2}+(n-1)(n-2)q^{2}}
    \nonumber \\
    &&+\frac{2(n-1)q^{2}}{nr^{2n-4}} \ _{2}F_{1}[\frac{n-2}{2(n-1)},\frac{1}{2},
    \frac{3n-4}{2(n-1)},
    -\frac{(n-1)(n-2)q^{2}}{2\beta^{2}r^{2n-2}}].
\end{eqnarray}
The solution (\ref{4eq1}) is asymptotically de Sitter. When $k\ne
1$, the solution (\ref{4eq1}) generalizes the so-called
topological de Sitter solutions proposed in \cite{CMZ} to the
Einstein-Born-Infeld theory with a positive cosmological constant.
When $k=1$ and $m>0$, in an appropriate parameter space, both
cosmological horizon $r_c$ and black hole horizon $r_+$ occur;
they satisfy $V(r)|_{r=r_+,r_c}=0$. When $k=0$ or $k=-1$, black
hole horizon disappears, but the cosmological horizon is still
there if $m<0$. As the case of asymptotic AdS, the black hole
horizon or cosmological horizon can be a positive, zero or
negative constant curvature hypersurface. Therefore these
solutions can be dubbed topologically asymptotic de Sitter
solutions.

According to the definition of mass due to Abbott and
Deser~\cite{ab}, the mass of the solution (\ref{4eq1}) is
\begin{equation}
M=(n-1)\omega_{n-1}m,
\end{equation}
where $m$ can be expressed in terms of black hole horizon radius
$r_+$ via $V(r_{+})=0$,
\begin{eqnarray}
\label{4eq3}
m &=& r_+^{n-2} \left ( k+ \left (\frac{4\beta^{2}}{n(n-1)}-\frac{1}{l^{2}}\right )r^{2}_{+}
  \right. \nonumber \\
&& -\frac{2\sqrt{2}\beta }{n(n-1)r_+^{n-3}}\sqrt{2\beta^{2}r^{2n-2}_{+}+(n-1)(n-2)q^{2}}
  \nonumber \\
&& \left. +\frac{2(n-1)q^{2}}{nr^{2(n-2)}_{+}} \
_{2}F_{1}[\frac{n-2}{2(n-1)},\frac{1}{2},\frac{3n-4}{2(n-1)},
-\frac{(n-1)(n-2)q^{2}}{2\beta^{2}r^{2n-2}_{+}}]\right ).
\end{eqnarray}
The Hawking temperature of black hole horizon is found to be
\begin{equation}
T=\frac{1}{4\pi r_+ }
   \left ((n-2) k+\left (\frac{4\beta^{2}}{n-1}-\frac{n}{l^{2}}\right
   )r_{+}^2
 -\frac{2\sqrt{2}\beta}{(n-1)r^{n-3}_{+}}\sqrt{2\beta^{2}r^{2n-2}_{+}+(n-1)(n-2)q^{2}}\right).
\end{equation}
The entropy of black hole horizon and the chemical potential
conjugate to the charge $Q$ are still given by (\ref{3eq3}) and
(\ref{3eq5}), respectively. Namely they keep the same forms as in
the case of asymptotically AdS space. In this way it is easy to
check that the first law of thermodynamics holds for the black
hole horizon, $dM=TdS+\Phi dQ$.

For the black hole horizon, we find that  the Casimir energy
$E_{c}$ is given b\cite{Cai2}
\begin{equation}
E_{c}\displaystyle=nE-(n-1)TS-(n-1)\Phi
Q=2k(n-1)r^{n-2}_{+}\omega_{n-1},
\end{equation}
and that the energy $E_{q}$ of the electromagnetic field outside
the black hole horizon remains the same form as the one given in
(\ref{3eq14}). Here the total energy $E=M$.

Then the entropy of black hole horizon given by (\ref{3eq3}) can
be recast to
\begin{equation}
\label{4eq6}
 S=\frac{2\pi
l}{n-1}\sqrt{E_{c}/k |(2(E-E_{q})-E_{c})|}.
\end{equation}
It is interesting to note that for the black hole in dS space, the
extensive part of total energy
\begin{equation}
2(E-E_q)-E_c =-2(n-1)\omega_{n-1}r_+^n/l^2,
\end{equation}
is negative. Its implication is not clear to us at the moment.

 Next we discuss the thermodynamics of cosmological horizon $r_c$.
 We relate the gravitational mass $\tilde M$ to the cosmological
 horizon $r_c$ by using the surface counterterm
 method~\cite{bbm,CMZ}. In this method, the gravitational mass is
 just the AD mass, but with an opposite sign,
\begin{equation}
\label{4eq8}
 \tilde{M}=-(n-1)\omega_{n-1}m,
\end{equation}
where $m$ can be expressed in terms of the cosmological horizon
radius $r_c$
\begin{eqnarray}
 m &=& r_c^{n-2} \left ( k+ \left (\frac{4\beta^{2}}{n(n-1)}-\frac{1}{l^{2}}\right )r^{2}_{c}
  \right. \nonumber \\
&& -\frac{2\sqrt{2}\beta
}{n(n-1)r_c^{n-3}}\sqrt{2\beta^{2}r^{2n-2}_{c}+(n-1)(n-2)q^{2}}
  \nonumber \\
&& \left. +\frac{2(n-1)q^{2}}{nr^{2(n-2)}_{c}} \
_{2}F_{1}[\frac{n-2}{2(n-1)},\frac{1}{2},\frac{3n-4}{2(n-1)},
-\frac{(n-1)(n-2)q^{2}}{2\beta^{2}r^{2n-2}_{c}}]\right ).
\end{eqnarray}
Note that the gravitational mass $\tilde M$ (\ref{4eq8}) is
measured on the boundary of the asymptotically dS spacetime,
namely at the future infinity (${\cal I}^+$), which is outside the
cosmological horizon. Similar to the black hole horizon, we can
obtain the Hawking temperature of the cosmological horizon
\begin{equation}
\tilde{T}=\frac{1}{4\pi r_c}\left (
    -(n-2)k -\left (\frac{4\beta^{2}}{n-1}-\frac{n}{l^{2}}\right )r_{c}^2
    + \frac{2\sqrt{2}\beta}{(n-1)r^{n-3}_{c}}\sqrt{2\beta^{2}r^{2n-2}_{c}+(n-1)(n-2)q^{2}}
    \right).
\end{equation}
In general, the Hawking temperature of cosmological horizon is not
equal to that of black hole horizon. Therefore the spacetime
describes a black hole in de Sitter space is unstable quantum
mechanically.  The chemical potential conjugate to the charge $Q$
is defined as
\begin{equation}
\tilde{\Phi}=-\sqrt{\frac{n-1}{2(n-2)}}\frac{16\pi q}{r^{n-2}_{c}}
\ _{2}F_{1}[\frac{n-2}{2n-2},\frac{1}{2},
\frac{3n-4}{2n-2},-\frac{(n-1)(n-2)q^{2}}{2\beta^{2}r^{2n-2}_{c}}].
\end{equation}
And the entropy of the cosmological horizon obeys the area formula
\begin{equation}
\tilde{S}=4\pi\omega_{n-1}r^{n-1}_{c}.
\end{equation}

For such defined thermodynamic quantities associated with the
cosmological horizon, we can see that they indeed satisfy the
first law of thermodynamics
\begin{equation}
d\tilde{M}=\tilde{T}d\tilde{S}+\tilde{\Phi}dQ.
\end{equation}
Therefore it is seemingly reasonable to consider black hole
horizon and cosmological horizon as two separate thermodynamic
systems~\cite{dS}.

For the cosmological horizon, the Casimir energy turns out to
be~\cite{Cai2,Cai3}
\begin{equation}
\tilde{E_{c}}\displaystyle=n\tilde{E}-(n-1)\tilde{T}\tilde{S}-(n-1)\tilde{\Phi}Q
   =-2k(n-1)r^{n-2}_{c}\omega_{n-1},
\end{equation}
where $\tilde E= \tilde M$. Similarly we can calculate the energy
of Born-Infeld electromagnetic field outside the cosmological
horizon
\begin{eqnarray}
\tilde E_{q} & \equiv & -\omega_{n-1} \int^{\infty}_{r_{c}}\bar
   T^{0}_{0}r^{n-1} dr \nonumber \\
 &=&- \omega_{n-1} \left ( \frac{4\beta^{2}}{n}r^{n}_{c}-\frac{2\sqrt{2}\beta r_{c}}
   {n}\sqrt{2\beta^{2}
 r^{2n-2}_{c}+(n-1)(n-2)q^{2}} \right. \nonumber \\
&& \left.
  +\frac{2(n-1)^{2}q^{2}}{nr^{n-2}_{c}} \ _{2}F_{1}[\frac{n-2}{2(n-1)},\frac{1}{2},
  \frac{3n-4}{2(n-1)},
 -\frac{(n-1)(n-2)q^{2}}{2\beta^{2}r^{2n-2}_{c}}]\right) .
\end{eqnarray}
Then we find that as the case of black hole horizon, the entropy
of cosmological horizon can be reexpressed in terms of the
Cardy-Verlinde formula
\begin{equation}
\tilde{S}=\frac{2\pi l}
   {n-1}\sqrt{|\tilde{E}_{c}/k|(2(\tilde{E}-\tilde{E_{q}})-\tilde{E_{c}})}.
\end{equation}
Here the extensive part of the energy
\begin{equation}
2(\tilde{E}-\tilde{E_{q}})-\tilde{E}_{c}=2(n-1)\omega_{n-1}r_c^2/l^2.
\end{equation}
Thus we show that both the entropies associated with black hole
horizon and cosmological horizon of the Born-Infeld black holes in
dS space can be formally recast to a form of the Cardy-Verlinde
formula, although the Casimir energy is negative in some cases, or
the extensive part of energy is negative in others. These results
imply that thermodynamics associated with black hole
(cosmological) horizon is indeed related to that of a some CFT.

\section{Factorized solution}
 In Sec.~II we mentioned that for the static electromagnetic field solution
 (\ref{2eq10}), we have $R''(r)=0$ from the Einstein equations of motion,
 and only the solution of $R=r$ was considered there, which gives
 the Born-Infeld black hole solutions in (A)dS spaces
 (\ref{2eq13}). In this section we discuss the case $R=a$ with $a$ being
 a positive constant. In this case, the metric (\ref{2eq6}) is a direct product
 of two subspaces, in which $ds_2^2= a^2 h_{ij}dx^idx^j$ is an
 ($n-1$)-dimensional constant curvature space.

 From the equation (\ref{2eq4}) of the Born-Infeld electromagnetic
 field, we have
 \begin{equation}
\label{5eq1}
 F^{rt}=\frac{\sqrt{(n-1)(n-2)}\beta
q}{\sqrt{2\beta^{2}a^{2n-2}+(n-1)(n-2)q^{2}}} \equiv f,
\end{equation}
where $q$ is still an integration constant, which can be related
to the electric charge of the solution via (\ref{2eq11}). Note
that here the electric field (\ref{5eq1}) is a constant,
independent of the radial coordinate $r$.

Solving (\ref{2eq7}) or (\ref{2eq8}) gives
\begin{equation}
V = -\Lambda_0 r^2 + c_1r +c_0,
\end{equation}
where $c_1$ and $c_0$ are two integration constants. Without loss
of generality, we can take $c_1=0$ through redefining the radial
coordinate $r$. And $c_0$ can be normalized to be one by rescaling
the coordinates $t$ and $r$. Then we have
\begin{equation}
V=1-\Lambda_0 r^2.
\end{equation}
 The constant $\Lambda_0$ is
\begin{equation}
\Lambda_0 = \frac{8}{n-1}\Lambda
-\frac{16\beta^2}{n-1}(1-\sqrt{1-f^2/\beta^2})-
\frac{8(n-3)f^2}{(n-1)\sqrt{1-f^2/\beta^2}}.
\end{equation}
From (\ref{2eq9}) we obtain a constraint
\begin{equation}
\label{5eq5}
 \frac{n-2}{a^2}k -\frac{2}{n-1}\Lambda
+\frac{4\beta^2}{n-1}
-\frac{4\beta^2}{(n-1)\sqrt{1-f^2/\beta^2}}=0.
\end{equation}
Therefore we see that the coordinates $t$ and $r$ in (\ref{2eq6})
now span a two-dimensional constant curvature spacetime. The
curvature is determined by $\Lambda_0$ (dS if $\Lambda_0>0$, AdS
if $\Lambda_0 <0$). In addition, once the parameters, the
curvature parameter $k$, charge $q$, the cosmological constant
$\Lambda$, and the Born-Infeld parameter $\beta$ are given, then
the curvature radius $a$ of the constant curvature space $ds_2^2 =
a^2 h_{ij}dx^idx^j$ can be determined via the relation
(\ref{5eq5}). Finally we obtain an exact solution (\ref{2eq6}),
which is a direct product of a two-dimensional dS (if
$\Lambda_0>0$) or AdS (if $\Lambda_0 <0$) space and an
($n-1$)-dimensional constant curvature space with a curvature
radius $a$.

\section{Conclusion}

We have found black holes solutions in the Einstein-Born-Infeld
theory with a cosmological constant in higher dimensions. These
black hole solutions are asymptotically de Sitter or anti-de
Sitter depending on the cosmological constant. Black hole horizon
or cosmological horizon in these solutions can be a positive, zero
or negative constant curvature hypersurface. Therefore these
solutions generalize some exact solutions of the
Einstein-Born-Infeld theory existing in the literature. We have
studied thermodynamics of black hole horizon and of cosmological
horizon.  For the Born-Infeld black holes in AdS spaces, when the
horizon is a zero or negative constant curvature hypersurface,
they are always thermodynamical stable with positive heat
capacity; when the horizon is a positive constant curvature
hypersurface, there is a critical value on the Born-Infeld
parameter for a fixed charge, above which the black hole is also
thermodynamical stable. However, the Born-Infeld parameter is less
than the critical value, a unstable black hole phase will appear.
In addition, it has been found that although the electromagnetics
of the Born-Infeld theory is non-linear, the entropies of black
hole horizon and cosmological horizon can all be recast to a form
of the Cardy-Verlinde formula, which indicates that the
Cardy-Verlinde formula is of universality in some sense. Note that
if there are some higher derivative terms of curvature, black hole
entropy in these gravitational theories can not be expressed in
terms of the Cardy-Verlinde formula~\cite{Cai}. Therefore our
results implies that higher order derivative terms of
electromagnetic field do not destroy the application of the
Cardy-Verlinde formula. In other words, the spacetime of the
Born-Infeld black hole at least in AdS space is still dual to a
some CFT residing on the boundary of the AdS space. The conformal
invariance of the dual CFT is not broken.

In the Einstein-Born-Infeld theory with a cosmological constant in
$D (\ge 4)$ dimensions, we have also found a factorized solution,
whose spacetime is a direct product of two constant curvature
spaces: one is a two-dimensional dS or AdS space and the other is
a $(D-2)$-dimensional positive, zero or negative constant
curvature space with a constant curvature radius.

\section*{Acknowledgements}

DWP thanks Hong-Sheng Zhang, Zong-Kuan Guo, Wei He, Jian-Huang
She, Qi Guo and Hao Wei for useful discussions and kind help. RGC
would like to express his gratitude to the Physics Department,
Baylor University for its hospitality. This work was supported by
Baylor University, a grant from Chinese Academy of Sciences, a
grant from NSFC, China (No. 13325525), and a grant from the
Ministry of Science and Technology of China (No. TG1999075401).


\begin{references}

\bibitem{topo} J.~P.~S.~Lemos,
                   Phys.\ Lett.\ B {\bf 353}, 46 (1995) [arXiv:gr-qc/9404041];
                  J.~P.~Lemos,
                   Class.\ Quant.\ Grav.\  {\bf 12}, 1081 (1995) [gr-qc/9407024];
                   J.~P.~S.~Lemos and V.~T.~Zanchin,
                   Phys.\ Rev.\ D {\bf 54}, 3840 (1996)
                   [arXiv:hep-th/9511188];
                   C.~G.~Huang and C.~B.~Liang,
                   Phys.\ Lett.\ A {\bf 201} (1995) 27;
                   R.~G.~Cai and Y.~Z.~Zhang,
                   Phys.\ Rev.\ D {\bf 54}, 4891 (1996) [arXiv:gr-qc/9609065];
                   R.~G.~Cai, J.~Y.~Ji and K.~S.~Soh,
                   Phys.\ Rev.\ D {\bf 57}, 6547 (1998)
                   [arXiv:gr-qc/9708063];
                   R.~G.~Cai,
                   Nucl.\ Phys.\ B {\bf 524}, 639 (1998) [arXiv:gr-qc/9801098];
                   S.~Aminneborg, I.~Bengtsson, S.~Holst and P.~Peldan,
                   Class.\ Quant.\ Grav.\  {\bf 13}, 2707 (1996)
                   [gr-qc/9604005];
                   R.~B.~Mann,
                   Class.\ Quant.\ Grav.\  {\bf 14}, L109 (1997) [gr-qc/9607071];
                   R.~B.~Mann,
                   Nucl.\ Phys.\ B {\bf 516}, 357 (1998)
                   [hep-th/9705223];
                   L.~Vanzo,
                   Phys.\ Rev.\ D {\bf 56}, 6475 (1997)
                   [gr-qc/9705004];
               D.~R.~Brill, J.~Louko and P.~Peldan,
               Phys.\ Rev.\ D {\bf 56}, 3600 (1997) [arXiv:gr-qc/9705012];
                  D.~Klemm,
                   Class.\ Quant.\ Grav.\  {\bf 15}, 3195 (1998) [gr-qc/9808051];
                   D.~Klemm, V.~Moretti and L.~Vanzo,
                   Phys.\ Rev.\ D {\bf 57}, 6127 (1998) [Erratum-ibid.\ D {\bf 60},
                   109902 (1998)] [gr-qc/9710123];
                   M.~Banados, A.~Gomberoff and C.~Martinez,
                   Class.\ Quant.\ Grav.\  {\bf 15}, 3575 (1998)
                   [hep-th/9805087];
        D.~Birmingham,
               Class.\ Quant.\ Grav.\  {\bf 16}, 1197 (1999)
               [hep-th/9808032];
        R.~Cai and K.~Soh,
              Phys.\ Rev.\ D {\bf 59}, 044013 (1999) [gr-qc/9808067];
K.~Behrndt, M.~Cvetic and W.~A.~Sabra,
Nucl.\ Phys.\ B {\bf 553}, 317 (1999) [arXiv:hep-th/9810227];
M.~Cvetic {\it et al.},
Nucl.\ Phys.\ B {\bf 558}, 96 (1999) [arXiv:hep-th/9903214];
M.~J.~Duff and J.~T.~Liu,
Nucl.\ Phys.\ B {\bf 554}, 237 (1999) [arXiv:hep-th/9901149];
                M.~F.~A.~da Silva, A.~Wang, F.~M.~Paiva and N.~O.~Santos,
                   Phys.\ Rev.\ D {\bf 61}, 044003 (2000) [arXiv:gr-qc/9911013];
                   R.~G.~Cai,
                   Phys.\ Lett.\ B {\bf 572}, 75 (2003)
                   [arXiv:hep-th/0306140];
R.~Aros, R.~Troncoso and J.~Zanelli,
Phys.\ Rev.\ D {\bf 63}, 084015 (2001) [hep-th/0011097];
R.~G.~Cai,
Phys.\ Rev.\ D {\bf 65}, 084014 (2002) [arXiv:hep-th/0109133];
R.~G.~Cai,
Phys.\ Lett.\ B {\bf 582}, 237 (2004) [arXiv:hep-th/0311240];
W.~L.~Smith and R.~B.~Mann,
Phys.\ Rev.\ D {\bf 56}, 4942 (1997) [gr-qc/9703007];
Y.~Wu, M.~F.~A.~da Silva, N.~O.~Santos and A.~Wang,
Phys.\ Rev.\ D {\bf 68}, 084012 (2003) [arXiv:gr-qc/0309002];
S.~Nojiri and S.~D.~Odintsov,
Phys.\ Lett.\ B {\bf 521}, 87 (2001) [Erratum-ibid.\ B {\bf 542},
301 (2002)] [arXiv:hep-th/0109122];
M.~Cvetic, S.~Nojiri and S.~D.~Odintsov,
Nucl.\ Phys.\ B {\bf 628}, 295 (2002) [arXiv:hep-th/0112045];
S.~Nojiri and S.~D.~Odintsov,
Phys.\ Rev.\ D {\bf 66}, 044012 (2002) [arXiv:hep-th/0204112];
Y.~M.~Cho and I.~P.~Neupane,
Phys.\ Rev.\ D {\bf 66}, 024044 (2002) [arXiv:hep-th/0202140];
I.~P.~Neupane,
Phys.\ Rev.\ D {\bf 67}, 061501 (2003) [arXiv:hep-th/0212092];
I.~P.~Neupane,
arXiv:hep-th/0302132;
R.~G.~Cai and A.~Wang,
arXiv:hep-th/0406040;
R.~G.~Cai and A.~z.~Wang,
Phys.\ Rev.\ D {\bf 70}, 064013 (2004) [arXiv:hep-th/0406057];
and references therein.

\bibitem{ab}L.~F.~Abbott and S.~Deser,
Nucl.\ Phys.\ B {\bf 195}, 76 (1982).

\bibitem{bbm}V.~Balasubramanian, J.~de Boer and D.~Minic,
Phys.\ Rev.\ D {\bf 65}, 123508 (2002) [arXiv:hep-th/0110108];
A.~M.~Ghezelbash and R.~B.~Mann,
JHEP {\bf 0201}, 005 (2002) [arXiv:hep-th/0111217];
D.~Klemm,
Nucl.\ Phys.\ B {\bf 625}, 295 (2002) [arXiv:hep-th/0106247].

\bibitem{CMZ}R.~G.~Cai, Y.~S.~Myung and Y.~Z.~Zhang,
Phys.\ Rev.\ D {\bf 65}, 084019 (2002) [arXiv:hep-th/0110234].
\bibitem{AdS}J.~Maldacena,
Adv.\ Theor.\ Math.\ Phys.\  {\bf 2}, 231 (1998) [Int.\ J.\
Theor.\ Phys.\  {\bf 38}, 1113 (1998)] [hep-th/9711200];
 S.~S.~Gubser, I.~R.~Klebanov and A.~M.~Polyakov,
Phys.\ Lett.\ B {\bf 428}, 105 (1998) [hep-th/9802109];
 E.~Witten,
Adv.\ Theor.\ Math.\ Phys.\  {\bf 2}, 253 (1998) [hep-th/9802150].

\bibitem{witten}E.~Witten,
Adv.\ Theor.\ Math.\ Phys.\  {\bf 2}, 505 (1998)
[arXiv:hep-th/9803131].

\bibitem{Stro}A.~Strominger,
JHEP {\bf 0110}, 034 (2001) [arXiv:hep-th/0106113].

\bibitem{Cardy}J.~L.~Cardy,
Nucl.\ Phys.\ B {\bf 270}, 186 (1986).

\bibitem{Ver}
E.~Verlinde,
arXiv:hep-th/0008140.

\bibitem{KPS}
 D.~Klemm, A.~C.~Petkou and G.~Siopsis,
Nucl.\ Phys.\ B {\bf 601}, 380 (2001) [arXiv:hep-th/0101076].

\bibitem{Cai}
 R.~G.~Cai,
Phys.\ Rev.\ D {\bf 63}, 124018 (2001) [arXiv:hep-th/0102113];
R.~G.~Cai, Y.~S.~Myung and N.~Ohta,
Class.\ Quant.\ Grav.\  {\bf 18}, 5429 (2001)
[arXiv:hep-th/0105070];
R.~G.~Cai and A.~z.~Wang,
Phys.\ Rev.\ D {\bf 70}, 064013 (2004) [arXiv:hep-th/0406057].

\bibitem{Birm}
 D.~Birmingham and S.~Mokhtari,
Phys.\ Lett.\ B {\bf 508}, 365 (2001) [arXiv:hep-th/0103108].
\bibitem{Jing}
J.~l.~Jing,
Phys.\ Rev.\ D {\bf 66}, 024002 (2002) [arXiv:hep-th/0201247].

\bibitem{Muck}L.~Cappiello and W.~Muck,
Phys.\ Lett.\ B {\bf 522}, 139 (2001) [arXiv:hep-th/0107238];
S.~Nojiri, S.~D.~Odintsov and S.~Ogushi,
Int.\ J.\ Mod.\ Phys.\ A {\bf 16}, 5085 (2001)
[arXiv:hep-th/0105117];
S.~Nojiri, S.~D.~Odintsov and S.~Ogushi,
Phys.\ Rev.\ D {\bf 65}, 023521 (2002) [arXiv:hep-th/0108172];
S.~Nojiri, S.~D.~Odintsov and S.~Ogushi,
Int.\ J.\ Mod.\ Phys.\ A {\bf 17}, 4809 (2002)
[arXiv:hep-th/0205187];
D.~Astefanesei, R.~Mann and E.~Radu,
JHEP {\bf 0401}, 029 (2004) [arXiv:hep-th/0310273].


\bibitem{Cai2}R.~G.~Cai,
Nucl.\ Phys.\ B {\bf 628}, 375 (2002) [arXiv:hep-th/0112253].

\bibitem{Cai3}R.~G.~Cai,
Phys.\ Lett.\ B {\bf 525}, 331 (2002) [arXiv:hep-th/0111093].

\bibitem{BI}
M.~Born and L.~Infeld,
Proc.\ Roy.\ Soc.\ Lond.\ A {\bf 144}, 425 (1934).

\bibitem{BIBH}D.~L.~Wiltshire,
Phys.\ Rev.\ D {\bf 38}, 2445 (1988);
D.~A.~Rasheed,
arXiv:hep-th/9702087.
T.~Tamaki and T.~Torii,
Phys.\ Rev.\ D {\bf 62}, 061501 (2000) [arXiv:gr-qc/0004071];
N.~Breton,
Phys.\ Rev.\ D {\bf 67}, 124004 (2003) [arXiv:hep-th/0301254];
M.~Aiello, R.~Ferraro and G.~Giribet,
arXiv:gr-qc/0408078.
M.~Cataldo and A.~Garcia,
Phys.\ Lett.\ B {\bf 456}, 28 (1999) [arXiv:hep-th/9903257];
S.~Fernando and D.~Krug,
Gen.\ Rel.\ Grav.\  {\bf 35}, 129 (2003) [arXiv:hep-th/0306120].

\bibitem{Dey}
T.~K.~Dey,
Phys.\ Lett.\ B {\bf 595}, 484 (2004) [arXiv:hep-th/0406169].

\bibitem{dS}C.~Teitelboim,
arXiv:hep-th/0203258;
A.~Gomberoff and C.~Teitelboim,
Phys.\ Rev.\ D {\bf 67}, 104024 (2003) [arXiv:hep-th/0302204];
R.~G.~Cai and Q.~Guo,
Phys.\ Rev.\ D {\bf 69}, 104025 (2004) [arXiv:hep-th/0311020].


\end{references}
\end{document}